\def\be{\begin{equation}}
\def\ee{\end{equation}}
\def\bea{\begin{eqnarray}}
\def\eea{\end{eqnarray}}
\def\bse{\begin{subequations}}
\def\ese{\end{subequations}}

\documentclass[prl,twocolumn,showpacs,amsmath,amssymb,preprintnumbers]{revtex4}
\usepackage{graphicx}
\usepackage{dcolumn}
\usepackage{bm}
\begin{document}
\title{Pre-asymptotic critical behavior and effective exponents in disordered metallic quantum ferromagnets}
\author{T.R. Kirkpatrick$^1$ and D. Belitz$^{2,3}$}
\affiliation{$^{1}$ Institute for Physical Science and Technology,
                    and Department of Physics, University of Maryland, College Park,
                    MD 20742, USA\\
                    $^{2}$ Department of Physics and Institute of Theoretical Science,
                    University of Oregon, Eugene, OR 97403, USA\\
                   $^{3}$ Materials Science Institute, University of Oregon, Eugene,
                    OR 97403, USA
            }
\date{\today}

\begin{abstract}
We determine the pre-asymptotic critical behavior at the quantum ferromagnetic transition in
strongly disordered metals. We find that it is given by effective power laws, in contrast to the
previously analyzed asymptotic critical behavior, which is valid only in an unobservably small
region. The consequences for analyzing experiments are discussed, in particular ways to
distinguish between critical behavior and Griffiths-phase effects.
\end{abstract}


\maketitle

The ferromagnetic quantum phase transition in metals has received a lot of interest in recent years.
In clean systems it is now well established that the transition is generically first order. 
This has been observed experimentally for ferromagnets as diverse as the d-electron metals 
MnSi and ZrZn$_2$ \cite{Pfleiderer_et_al_1997, Uhlarz_Pfleiderer_Hayden_2004}, and
various uranium-based compounds \cite{Aoki_et_al_2011a}. It is in sharp contrast to early theories that 
predicted a second-order transition with mean-field exponents for both clean and disordered quantum 
ferromagnets \cite{Hertz_1976}, but in excellent agreement with a theory that takes into account the 
coupling between the magnetization and soft or gapless fermionic excitations in metals 
\cite{Belitz_Kirkpatrick_Vojta_1999, Kirkpatrick_Belitz_2012b}. Disordered systems, which include
almost all magnets where the quantum phase transition is triggered by chemical composition, are not
as well understood. Experimentally, continuous or second-order transitions have been observed in a
variety of systems, but attempts to determine exponents have yielded very different
results for different systems, and even for different analyses of the same 
system \cite{Bauer_et_al_2005, Butch_Maple_2009, Ubaid-Kassis_Schroeder_2008,
Ubaid-Kassis_Vojta_Schroeder_2010}. Theoretically, the same framework that predicts a 
first-order transition in clean systems predicts a second-order one in the presence of 
quenched disorder \cite{Kirkpatrick_Belitz_1996, Belitz_et_al_2001a, Belitz_et_al_2001b}. Again, the coupling
of gapless excitations (in this case, diffusive particle-hole excitations) leads to strong deviations from
the prediction of Hertz theory \cite{strong_disorder_footnote}. The asymptotic critical behavior is 
given by non-mean-field power laws modified by multiplicative log-normal terms. Comparisons
between this asymptotic critical behavior and experimental results are overall not satisfactory.
Another important development has been the study of quantum Griffiths effects that are expected to 
be present in systems with quenched disorder \cite{Millis_Morr_Schmalian_2002, Vojta_2010}. These 
are due to rare regions devoid of disorder, occur in a
whole region in the phase diagram,
and are characterized by non-universal power laws.
The interplay between Griffiths-phase effects and critical phenomena is incompletely understood and
makes the interpretation of experiments difficult, see the discussion at the end. 
 
The quantum critical point is difficult to determine
experimentally, and the distance from it can be hard to control. As a result, the experimental situation
is much more difficult than in the case of classical phase transitions. For the latter, it is known that
in almost all cases an observation of the asymptotic critical behavior requires measurements over
two decades or more in a region within less (in some cases substantially less) than 1\%
from the critical point \cite{Levelt-Sengers_Sengers_1981}. Such precision is currently 
not achievable for quantum phase
transitions. Further, for classical transitions it is known that the asymptotic critical region is
preceded by a pre-asymptotic region where observables obey apparent power-law behavior with
effective exponents that can be very different from the asymptotic ones. 
 
It is the purpose of this Letter to analyze the corresponding pre-asymptotic behavior for the 
quantum ferromagnetic transition in disordered metals. This is a necessary step towards a better 
understanding of experimental results on such systems. We find that for strongly disordered 
systems \cite{strong_disorder_footnote} the pre-asymptotic region (which we estimate to range from 
about 0.1\% to 10\% away from the critical point) is governed by effective power laws that can be 
represented by homogeneity laws for the magnetization $m$, the spin susceptibility $\chi$, the specific-heat 
coefficient $\gamma$, and the electrical conductivity $\sigma$:
\bse
\label{eqs:1}
\bea
m(t,T,h) &=& b^{-2}\,f_m(t\,b^{d-2+\lambda}, T\,b^{2+\lambda}, h\,b^{d+\lambda})\ ,
\label{eq:1a}\\
\chi(t,T,h) &=& b^{d-2+\lambda}\,f_{\chi}(t\,b^{d-2+\lambda}, T\,b^{2+\lambda}, h\,b^{d+\lambda})\ ,\qquad
\label{eq:1b}\\
\gamma(t,T,h) &=& b^{\lambda}\,f_{\gamma}(t\,b^{d-2+\lambda}, T\,b^{d+\lambda}, h\,b^{d+\lambda})\ ,
\label{eq:1c}\\
\sigma(t,T,\omega) &=& {\rm const.} + b^{-(d-2)}\,f_{\sigma}(t\,b^{d-2+\lambda}, T\,b^{d+\lambda}, 
\nonumber\\
&&\hskip 100pt \omega\,b^{d+\lambda})\ .\quad
\label{eq:1d}
\eea
\ese
Here the $f$ are scaling functions, $d$ is the dimensionality, $t$ is the dimensionless distance from 
the quantum critical point, and $T$, $h$, and $\omega$ denote the
temperature, magnetic field, and frequency, respectively. $b$ is the arbitrary renormalization-group scale factor. 
$\lambda$ is an effective exponent that depends on $d$. Strictly speaking, it varies with $t$ and approaches zero
for $t\to 0$; however, this happens only for unobservably small values of $t$. For the $t$-range
mentioned above, which is the relevant one for all existing experiments, our numerical results (see below)
show
\be
\lambda \approx 2/3\ .
\label{eq:2}
\ee
Note that the scale dimension of the temperature depends on the observable considered: As an argument of
$m$ or $\chi$ it is $2 + \lambda$, but as an argument of $\gamma$ or $\sigma$,
$d + \lambda$. This is because of the presence of two time scales in the system; a critical one 
and and a renormalized diffusive one with dynamical exponents $z_c = d + \lambda$ and 
$z_d = 2 + \lambda$, respectively. Determining which time scale controls a given observable
requires detailed calculations \cite{Belitz_et_al_2001b}. Suitable choices of $b$ yield the predicted 
power-law behavior of the observables in the pre-asymptotic
region. For instance, in $d=3$ we obtain $m(t,0,0) \propto \vert t\vert^{\beta}$,
$m(0,0,h) \propto h^{1/\delta}$, and $m(0,T,h) = T^{\beta_T} f_m(0,1,h/T^{\beta_T \delta})$ with effective critical
exponents
\bse
\label{eqs:3}
\bea
\beta &=& 2/(1 + \lambda) \approx 6/5\quad,\quad 
\beta_T = 2/(2 + \lambda) \approx 3/4\quad,\quad
\nonumber\\
\delta &=& (3 + \lambda)/2 \approx 11/6\ .
\label{eq:3a}
\eea
Similarly, we have $\gamma(t,0,0) \propto \vert t\vert^{-\kappa}$
and $\chi(t,0,0) \propto \vert t\vert ^{-\gamma}$, with effective exponents
\be
\kappa = \lambda/(1 + \lambda) \approx 2/5\quad,\quad \gamma = 1\ .
\label{eq:3b}
\ee
The corresponding temperature scaling at criticality is $\gamma(0,T,0) \propto T^{-{\kappa_T}}$ and
$\chi(0,T,0) \propto T^{-\gamma_T}$, with
\be
\kappa_T = \lambda/(1 + \lambda) \approx 2/11\quad,\quad \gamma_T = (1+\lambda)/(2+\lambda) \approx 5/8\ .
\label{eq:3c}
\ee
Note that 
the theory predicts $h/T$ scaling for $\gamma$, but not for $m$ and $\chi$.
Also of interest is the electrical resistivity $\rho = 1/\sigma$. For the critical 
contribution $\delta\rho$ Eq.\ (\ref{eq:1d}) predicts $\delta\rho(t=0,T) \propto T^{s_T}$ with
\be
s_T = 1/(3 + \lambda) \approx 3/11\ ,
\label{eq:3d}
\ee
\ese 
and $\delta\rho$ obeys $\omega/T$ scaling.
  
To derive these results we go back to the exact theory for the quantum critical point in 
disordered metallic ferromagnets that was
developed in Refs.\ \cite{Belitz_et_al_2001a, Belitz_et_al_2001b}. It takes the form of a
coupled field theory for the order-parameter fluctuations and the diffusive fermionic excitations
characteristic of disordered metals. The former are characterized by a two-point vertex $u_2$
which, in the bare theory, takes the usual Ornstein-Zernike form as a function of the wave number ${\bm k}$,
\be
u_2({\bm k}) = t_0 + a_2{\bm k}^2\ ,
\label{eq:4}
\ee
with $t_0$ the bare distance from the critical point and $a_2$ a coefficient.  The inverse of $u_2$
gives the bare (i.e, Landau-Ginzburg) spin susceptibility $\chi({\bm k}) = 1/u_2({\bm k})$. The fermionic 
excitations are expressed in terms of a diffusive propagator ${\cal D}$. 
In the bare theory it has the form
\be
{\cal D}({\bm k},\Omega_n) = \frac{1}{{\bm k}^2 + GH\Omega_n}\ ,
\label{eq:5}
\ee
with $\Omega_n$ a bosonic Matsubara frequency.
Here $G$ is proportional to the disorder strength, $H$ 
is proportional to the inverse of the bare specific-heat coefficient $\gamma = C/T$, and $GH$ is the bare
heat diffusion coefficient. A coupling between the diffusive electronic excitations and the
order-parameter fluctuations mediated by a spin-triplet interaction constant $K_t$ leads to the 
well-known dynamics of the order-parameter fluctuations described by the paramagnon propagator
\cite{Doniach_Engelsberg_1966, Hertz_1976}
\be
{\cal P}({\bm k},\Omega_n) = \frac{1}{u_2({\bm k}) + \frac{GK_t\vert\Omega_n\vert}{{\bm k}^2 + GH\vert\Omega_n\vert}}\ .
\label{eq:6}
\ee

In Refs. \cite{Belitz_et_al_2001a, Belitz_et_al_2001b} the renormalizations of the above quantities
have been studied in the vicinity of the quantum critical point where the susceptibility
and the specific-heat coefficient diverge. An important effect that arises at one-loop order is the
appearance of a term $\propto\vert{\bm k}\vert^{d-2}$ in $u_2$. This term was
missing in Hertz's tree-level analysis \cite{Hertz_1976}. If it is included, one finds a Gaussian
fixed point that  leads to a standard power-law critical behavior with non-mean-field 
exponents \cite{Kirkpatrick_Belitz_1996}. This fixed point is marginally unstable due to the 
existence of various marginal operators. It was shown in Refs. \cite{Belitz_et_al_2001a, Belitz_et_al_2001b} 
that the effects of the latter can be determined exactly, and the critical behavior
can thus be determined exactly in all dimensions $d>2$. 
It can be described by coupled integral
equations for the spin diffusion coefficient $D_{\text{s}}({\bm k},\Omega_n)$ and the thermal diffusion
coefficient $D(\Omega_n)$. $D_{\text{s}}$ is proportional to the dressed order-parameter vertex 
$u_2({\bm k},\Omega_n)$, which acquires a frequency dependence under renormalization. $D$ is proportional
to the dressed coupling constant $H(\Omega_n)$; $G$ does not acquire any renormalizations that
lead to singular behavior at the critical point. The integral equations take the form \cite{Belitz_et_al_2001b}
\bse
\label{eqs:7}
\bea
D_{\text{s}}({\bm k},\Omega) &=& D_{\text{s}}^0 + \frac{iG}{2V}\sum_{\bm p} \int_0^{\infty} d\omega\,
   \frac{1}{{\bm p}^2 - i\omega/D(\omega)}
\nonumber\\
&&\times \frac{1}{({\bm p}+{\bm k})^2 - i(\omega + \Omega)/D(\omega + \Omega)}\ ,
\label{eq:7a}\\
\frac{1}{D(\Omega)} &=& \frac{1}{D^0} + \frac{3G}{8V} \sum_{\bm p} \frac{1}{\Omega} \int_0^{\Omega} d\omega\,
   \frac{1}{-i\omega + {\bm p}^2 D_{\text{s}}({\bm p},\omega)}\ ,
   \nonumber\\
\label{eq:7b}
\eea
\ese
with $D_{\text{s}}^0$ and $D^0$ the bare spin and heat diffusion coefficients. 
The solution of these equations gives the leading behavior of the spin susceptibility 
$\chi({\bm k},\Omega) \propto 1/D_{\text{s}}({\bm k},\Omega)$ and the specific-heat coefficient
$\gamma = 1/D(\Omega=0)$ in the vicinity of the critical point where $D_{\text{s}}(0,0) = D(0) = 0$.
They hold in the disordered phase; the behavior of the magnetization can be inferred from scaling
considerations \cite{Belitz_et_al_2001a, Belitz_et_al_2001b}. The temperature dependence of 
$\gamma$ and $\chi$ is the same as the frequency dependence on scaling grounds.

The Eqs. (\ref{eqs:7}) have an interesting history. They were first derived in the context of
the metal-insulator transition in disordered interacting electron systems \cite{Belitz_Kirkpatrick_1991,
Kirkpatrick_Belitz_1992}. They were solved by various methods, including a numerical solution and a
renormalization-group solution. The solutions showed that they describe a phase transition at a critical value
$G_c$ of $G$ where the spin dynamics freeze, but the nature of the frozen phase was unclear. Only later
was it realized that they describe the quantum ferromagnetic transition \cite{Kirkpatrick_Belitz_1996}. This
relation was fully explored in Refs.\ \cite{Belitz_et_al_2001a, Belitz_et_al_2001b}, which used
the asymptotic solution available from the earlier work. At criticality, the solution can be expressed in terms of
a function
\bse
\label{eqs:8}
\be
g(x) = \sum_{n=0}^{\infty} \frac{[c(d) x]^n}{n!}\,e^{(n^2-n)\ln(2/d)/2}\ ,
\label{eq:8a}
\ee
where $c(d)$ is a dimensionality-dependent constant that has a not been determined analytically. The leading
behavior of $g$ for large arguments is log-normal,
\be
g(x\to\infty) = \frac{1}{\sqrt{\pi}}\,e^{(\ln x)^2/2\ln(d/2)}\ .
\label{eq:8b}
\ee
\ese
$g$ determines the asymptotic critical behavior of various observables, which can be cast in the form of
homogeneity laws. For the specific-heat coefficient 
this takes the form
\be
\gamma(t,T) = g(\ln b)\,f_{\gamma}(t\,b^{d-2} g(\ln b), T\,b^d\,g(\ln b))\ ,
\label{eq:9}
\ee
with $f_{\gamma}$ a scaling function. 
Equation (\ref{eq:9}) implies in particular that at $T=0$, upon approaching criticality, $\gamma$ diverges
faster than any power, but slower than exponentially: $\gamma(t\to 0) \propto g(\ln(1/t))$. Other observables
obey corresponding homogeneity laws, and their asymptotic critical behavior is given by power laws multiplied
by the log-normal function represented by Eq.\ (\ref{eq:8b}). 

The analysis of Ref.\ \cite{Belitz_et_al_2001b}, while asymptotically exact, is not very useful for analyzing
experiments. The reason is that the asymptotic region is much smaller than what is realistically accessible
in experiments. Indeed, Eq.\ (\ref{eq:8b}) is valid only for $\ln x \gg 1$, and
the asymptotic critical region is therefore exponentially small. To determine the experimentally
relevant behavior we resort to a numerical solution of the integral equations. 
\begin{figure}[t]
\vskip -0mm
\includegraphics[width=8.5cm]{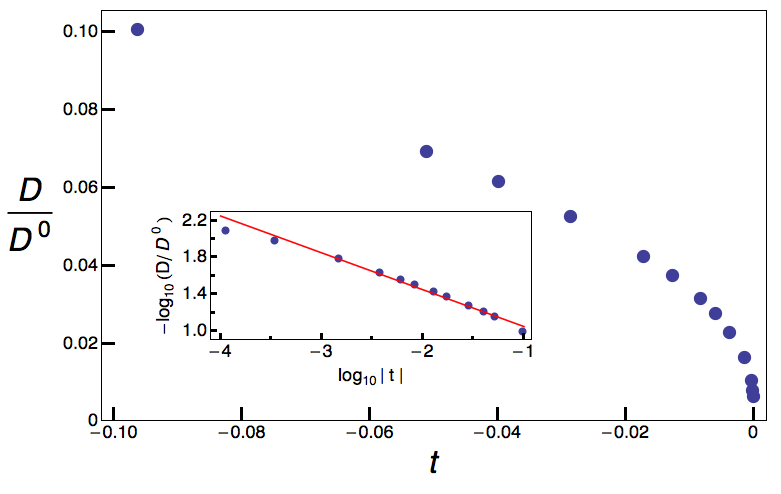}
\caption{(Color online)  Solution of Eqs.\ (\ref{eqs:7}). Main frame: 
              $D(\Omega=0)/D^0$ as function of $t$. Inset: log-log plot of the same data. The straight 
              line corresponds to a slope $\kappa = 0.4$.              }
               \vskip -0mm
\label{fig:1}
\end{figure}
We use a soft momentum cutoff and dimensionless units as in Ref.\ \cite{Belitz_Kirkpatrick_1991}. 
Figure \ref{fig:1} shows the normalized heat diffusion coefficient $D(\Omega=0)/D^0$ vs 
$t = (G - G_{\text{c}})/G_{\text{c}}$ for $d=3$. The inset shows that the behavior is very well described by
an effective power law $D(\Omega=0) \propto t^{-\kappa}$ with $\kappa = 0.4$. The only input are
the two Fermi-liquid parameters $F_0^a$ and $F_1^s$, which determine $D_{\text{s}}^0$. For
Fig.\ (\ref{fig:1}) we have chosen $F_0^a = -0.9$ and $F_1^s = 0$ as in Ref.\ \cite{Belitz_Kirkpatrick_1991};
this reproduces the results of that paper \cite{parameters_footnote}. The important
new point in the present context is that this effective power law implies, via Eq.\ (\ref{eq:9}), that 
$g(\ln b) \approx b^{\lambda}$ with $\lambda = \kappa/(1 - \kappa)$ in a range of $b$ values. This turns
Eq.\ (\ref{eq:9}) into Eq.\ (\ref{eq:1c}), and results in many other power laws.  As a check of the universality
of our scaling assumption, namely, that $g(\ln b)$ in the pre-asymptotic region effectively turns into $b^{\lambda}$
in {\em all} expressions it enters, we have calculated $D_{\text{s}}(0,\Omega)$ and $D(\Omega)$ at
criticality from the integral equations. 
\begin{figure}[t]
\vskip -0mm
\includegraphics[width=8.5cm]{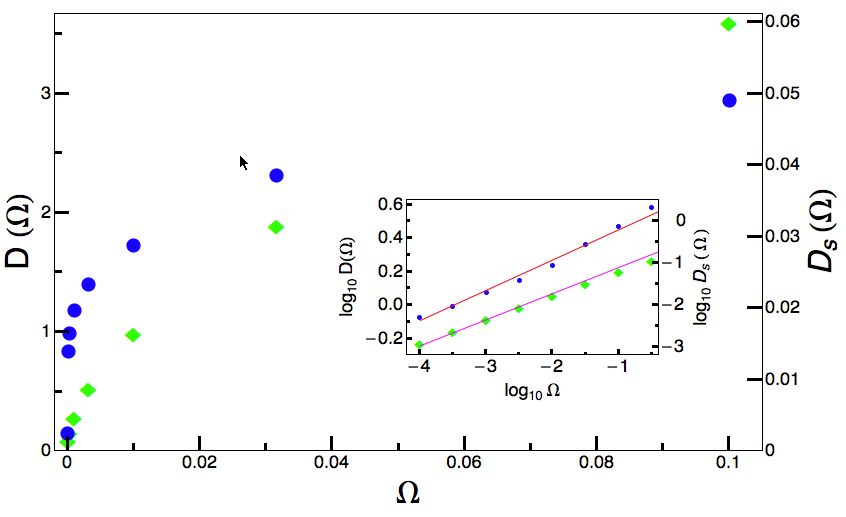}
\caption{(Color online) $D(\Omega)$ (blue dots) and $D_{\text{s}}(0,\Omega)$ (green diamonds) at criticality. 
               $\Omega$, $D$, and $D_{\text{s}}$ are measured in atomic units as in Ref.\ \cite{Belitz_Kirkpatrick_1991}.              
               Inset: log-log plot of the same data; the straight lines correspond to $\kappa_T = 0.181$
               and $\gamma_T = 0.625$.             
              }
               \vskip -0mm
\label{fig:2}
\end{figure}
Figure \ref{fig:2} shows that the numerical solution is in excellent agreement with the effective exponents
from Eq.\ (\ref{eq:3c}) over two decades of the frequency or temperature. This is a nontrivial check of the effective
scaling assumption, as two dynamical effective exponents ($\kappa_T$ and $\lambda_T$) are correctly predicted 
from the numerical determination of one static one ($\kappa$). As a consistency check we have also
calculated $g(x)$ as defined in Eq.\ (\ref{eq:8a}) numerically in the $x$-region that corresponds to the $t$-region
over which the effective exponent $\kappa$ is found. A necessary consistency requirement is that
$g(x) \approx \exp(\lambda x)$ in the appropriate region for some value of $c(d)$. We have found that this is
indeed the case for $c(3) \approx 1.4$. 

These results demonstrate that the observable ferromagnetic quantum critical behavior if given by the homogeneity
laws expressed in Eqs.\ (\ref{eqs:1}). The exponent $\lambda$ is strictly speaking not universal, i.e., it depends on
$t$, but for $t$ between about $10^{-3}$ and $10^{-1}$ it is given, in $d=3$, by Eq.\ (\ref{eq:2}) with an accuracy
as demonstrated by the log-log plots in Figs. \ \ref{fig:1} and \ref{fig:2}. Asymptotically close to the transition the 
behavior crosses over to the non-power-law asymptotic solutions discussed in Ref.\ \cite{Belitz_et_al_2001b},
but this happens only unobservably close to the critical point. 

We stress that this critical behavior is expected to be present {\em in addition} to any Griffiths-phase effects, and that
the two will be independent. This is because the fixed point found in Ref.\ \cite{Belitz_et_al_2001a}, which governs
both the asymptotic critical behavior and the pre-asymptotic one discussed above, satisfies the Harris
criterion \cite{Harris_1974, Chayes_et_al_1986}. Griffiths effects therefore cannot modify the critical behavior;
this is a case of coexisting critical behavior and quantum Griffiths singularities, as discussed in 
Refs.\ \cite{Vojta_Hoyos_2014, Vojta_Igo_Hoyos_2014}. We note that the critical behavior manifests
itself on either side of the transition, whereas Griffiths phase effects are present only in the disordered phase.

We close with remarks concerning the experimental relevance of our results. Bauer et al. \cite{Bauer_et_al_2005}
found a ferromagnetic quantum critical point in URu$_{2-x}$Re$_x$Si$_2$ at $x\approx 0.3$. Scaling plots of the
magnetization in an external field yielded exponents (in our notation) $\delta = 1.56$, $\beta_T = 0.9$, and
$\gamma_T = \beta_T\,(\delta - 1) = 0.5$ \cite{Widom_footnote}. Comparing with Eqs.\ (\ref{eqs:3}) we see that the 
theoretical pre-asymptotic exponents are within 20\% of these values. However, the temperature dependence of the
resistivity in the vicinity of the critical concentration was found to be roughly linear, and thus much weaker than
the prediction of Eq.\ (\ref{eq:3d}). A later analysis \cite{Butch_Maple_2009}
put the critical concentration at $x \approx 0.15$ and found continuously varying exponents in the range
$0.6 \geq x \geq 0.2$, with $\delta\to 1$, $\gamma_T\to 0$, and $\beta_T \approx 0.8$ roughly 
constant. These data were taken in what the authors interpreted as the ordered phase, so Griffiths-phase effects,
which might explain the continuously varying exponents, should not be present. If the data represent critical
phenomena, then continuously varying exponents are hard to understand, and an exponent $\gamma_T = 0$,
which must signify an order-parameter susceptibility that diverges only logarithmically, would be very
unusual.

Another relevant material is Ni$_{1-x}$V$_x$. Ref.\ \cite{Ubaid-Kassis_Schroeder_2008} found a critical point
at $x_c \approx 0.11$ with $\gamma_T = 0.37 \pm 0.07$, $\beta_T \approx 0.5$ and $\delta = 1.8 \pm 0.2$.
The value of $\delta$ agrees very
well with Eqs.\ (\ref{eqs:3}), the agreement for $\gamma_T$ and $\beta_T$ is less satisfactory. These data were
reinterpreted in Ref.\ \cite{Ubaid-Kassis_Vojta_Schroeder_2010} in terms of a Griffiths phase for
$x > x_c$. The interpretation of the susceptibility required adding a Curie component
that was attributed to the presence of frozen spin clusters. Our results predict that near the critical concentration
a third component should be present, which is proportional to $T^{-(1+\lambda)/(2+\lambda)}
\approx T^{-5/8}$, or $h^{-(1+\lambda)/(3+\lambda)} \approx h^{-5/11}$. This should be present in both
the ordered and the disordered phase, and the $T$ or $h$ window where it is valid will depend on the
distance from the critical point.

Finally, a value of $\delta \approx 1.5$ has been reported for the transition observed in Sr$_{1-x}$A$_x$RuO$_3$,
where the dopant A can be Ca, La$_{0.5}$Na$_{0.5}$, or La \cite{Itoh_Mizoguchi_Yoshimura_2008}. 

In summary, our results imply that a complete understanding of the ferromagnetic quantum phase transition
in disordered systems requires, (1) a careful analysis on both sides of the transition, (2) taking into account
the pre-asymptotic critical behavior, and (3) measurements of the magnetization, which is the only observable
for which critical behavior can be reliably separated from Griffiths-phase effects.

\acknowledgments
This work was supported by
the NSF under grant Nos. DMR-1401410 and DMR-1401449. Part of this work was performed at the
Aspen Center for Physics and supported by the NSF under Grant No. PHYS-1066293.


\begin{thebibliography}{31}
\bibitem[{par()}]{parameters_footnote}
\bibinfo{note}{Instead of changing $G$ at fixed $F_0^a$ one can trigger the
  transition by changing $F_0^a$ at fixed $G$, or any combination of the two.
  $G_c$ depends on $F_0^a$ and on the momentum cutoff, but the effective
  exponent $\kappa$ does not discernibly depend on these choices. More
  generally, one needs to start with bare values sufficiently close to the
  transition for the renormalization procedure to converge to the critical
  fixed point. Once this is the case, the asymptotic results are independent of
  the initial conditions, and the pre-asymptotic ones depend only weakly on
  them.}
\bibitem[{Wid()}]{Widom_footnote}
\bibinfo{note}{Note that Widom's scaling law, $\gamma = \beta (\delta -1)$,
  still holds if $\beta$ and $\gamma$ are replaced by $\beta_T$ and $\gamma_T$
  as defined in the context of Eqs.\ (\ref{eqs:3}).}
\end{thebibliography}

\end{document}